# Causality and Chance in Relativistic Quantum Field Theories

Richard A. Healey

*Abstract*

Bell appealed to the theory of relativity in formulating his principle of local causality. But he maintained that quantum field theories do not conform to that principle, even when their field equations are relativistically covariant and their observable algebras satisfy a relativistically motivated microcausality condition. A pragmatist view of quantum theory and an interventionist approach to causation prompt the reevaluation of local causality and microcausality. Local causality cannot be understood as a reasonable requirement on relativistic quantum field theories: it is unmotivated even if applicable to them. But microcausality emerges as a sufficient condition for the consistent application of a relativistic quantum field theory.

## 1 Introduction

The analysis of causal concepts and principles in relativistic quantum theories is often now pursued in the framework of algebraic quantum field theory. While this has permitted an increase in mathematical rigor, that alone has not produced a resolution of difficulties highlighted by Bell in his seminal examination of causality in what he called ordinary quantum mechanics. Bell appealed to the theory of relativity in formulating his principle of local causality. But he maintained that, even when their field equations are relativistically covariant and their observable algebras satisfy a relativistically motivated microcausality condition, quantum field theories do not conform to that principle. A pragmatist view of quantum theory and an interventionist approach to causation prompt the reevaluation of local causality and microcausality. I shall use these to argue that local causality cannot be understood as a reasonable requirement on relativistic quantum field theories. Microcausality, on the other hand, emerges as a sufficient condition for the consistent application of a relativistic quantum field theory.

The structure of the paper is as follows. I begin by reviewing Bell's argument that ordinary quantum mechanics is not locally causal and cannot be embedded in a locally causal theory. While recent work has helped to clarify the structure of this argument, it is still necessary to highlight its reliance on assumptions about the kind of theory to which it applies. In section 3 I sketch a pragmatist view of quantum theory I have advocated elsewhere and explain why it is not straightforward to apply Bell's condition of local causality to quantum theory, so viewed. The local causality condition is either inapplicable to quantum theory or unmotivated by the intuitive principle on which Bell based it.

Section 4 applies an interventionist approach to causation suggested by Bell's own remarks to argue that violation of Bell inequalities in an EPR-Bell scenario is no sign of superluminal causation. Counterfactual relations between space-like separated events supported by quantum correlations are not causal because they are not stable under hypothetical interventions.

After reluctantly abandoning local causality, Bell went on to treat with skepticism the suggestion that a no-superluminal-signalling requirement could serve to express the fundamental causal structure of theoretical physics, as implemented by a principle of microcausality. In section 5 I argue that a strong form of microcausality can be justified as a sufficient condition for the consistent assignment of states in ordinary relativistic quantum mechanics. While it does play a role in an argument against superluminal causation, a no-superluminal-signalling requirement does not express the fundamental causal structure of theoretical physics. This should disappoint no one. A precise formulation of physical theory will not use causal notions. Rather than inhering in the structure of theoretical models, causal structure emerges only when one adopts the perspective of a physically situated agent thinking of applying these models.

Section 6 briefly extends this analysis of local causality and microcausality to relativistic algebraic quantum field theory. While this does not substantially affect its conclusions, a standard formulation of microcausality in this context is already strong enough to guarantee a consistent, relativistically invariant assignment of states.

## 2 Local Causality

In "La Nouvelle Cuisine", the last of an important series of papers (reprinted in his [2004]), J.S. Bell argued that any seriously formulated theory meeting a local causality condition based on an intuitive conception of local action tailored to the structure of relativistic space-time predicts correlations different from those successfully predicted by quantum theory. Applying this condition to "ordinary quantum mechanics" he concluded that quantum theory is neither locally causal nor embeddable in a locally causal theory. He had already made clear in an earlier paper ([2004], p.55) that he took ordinary quantum mechanics to include relativistic quantum field theory. If this conclusion follows, then not only does relativistic quantum field theory conflict with the intuitive conception of relativistic local action, but so would any seriously formulated theory compatible with some of its experimentally verified predictions.

The dialectic of the previous paragraph has produced something of a consensus among those most closely influenced by Bell's writings that the observed correlations demonstrate the falsity of the intuitive principle—in that sense the world itself is non-local. So it is important to see whether Bell's conclusion does indeed follow, and to be clear on whether quantum theory satisfies the condition of local causality or can be squared with the intuitive conception on which Bell based it.

Bell ([2004], p.239) begins his argument by stating the following intuitive principle of local causality:

> The direct causes (and effects) of events are near by, and even the indirect causes (and effects) are no further away than permitted by
the velocity of light.

Appropriately, he takes this to be too imprecise to serve as a premise in a mathematical argument, involving as it does the vague notions of cause and effect. Such words as 'cause' and 'effect' will not appear in the formulation of any serious theory. But as he points out elsewhere this does not mean that one cannot investigate the causal structure of a theory:

> I would insist here on the distinction between analyzing various physical theories on the one hand, and philosophising about the unique real world on the other. In this matter of causality it is a great inconvenience that the world is given to us once only. We cannot know what would have happened if something had been different. ... Physical theories are more amenable in this respect. We can calculate the consequences of changing free elements in a theory, be they only initial conditions, and so explore the causal structure of the theory. ([2004], p.101)

Indeed, Bell ([2004], p.239) continues by formulating a local causality condition on physical theories of a certain kind based on his intuitive principle. In doing so he uses the important neologism 'beable', which he had previously introduced as follows:

> It is not easy to identify precisely which physical processes are to be given the status of "observations" and which are to be relegated to the limbo between one observation and another. So it could be hoped that some increase in precision might be possible by concentration on the beables, which can be described "in classical terms", because they are there. The beables must include the settings of switches and knobs on experimental equipment, the currents in coils, and the readings of instruments. "Observables" must be made, somehow, out of beables. (Bell [2004], p.52)

After this introduction, he had gone on to distinguish a class of local beables:

> We will be particularly concerned with local beables, those which (unlike the total energy) can be assigned to some bounded space-time region. ...
>
> It is in terms of local beables that we can hope to formulate some notion of local causality. (Bell [2004], p.53)

Section 3 of "La Nouvelle Cuisine" gives his reasons to hope.

> No one is obliged to consider the question "What cannot go faster than light?". But if you do decide to do so, then the above remarks suggest the following: you must identify in your theory 'local beables'. The beables of the theory are those entities in it that are,

> at least tentatively, to be taken seriously, as corresponding to something real. ([2004], p.234)

In "the above remarks" Bell had distinguished elements of a classical theory (such as the electric and magnetic fields of Maxwell's electromagnetism) intended to represent something real from elements (such as the electric and magnetic potentials) that function only as calculational tools, not to correspond to something real. Note that while Bell introduces beables as "entities", his examples can all be recast as (values of) physical magnitudes, respecting the parallel between beables and observables.

He continues

> Now it may well be that there just are no local beables in the most serious theories. When space-time itself is "quantized", as is generally held to be necessary, the concept of locality becomes very obscure. ... So all our considerations are restricted to that level of approximation to serious theories in which space-time can be regarded as given, and localization becomes meaningful. Even then, we are frustrated by the vagueness of contemporary quantum mechanics. You will hunt in vain in the text-books for the local beables of the theory. What you may find there are the so-called "local observables". It is then implicit that the apparatus of "observation", or better, of experimentation, and the experimental results, are real and localized. We will have to do as best we can with these rather ill-defined local beables, while hoping always for a more serious reformulation of quantum mechanics where the local beables are explicit and mathematical rather than implicit and vague. ([2004], p.235)

It will be important in what follows to distinguish between beables (local or otherwise) that figure in a serious formulation of a theory such as quantum theory and beables acknowledged in applications of that theory. The metric, Riemann and stress-energy tensors will figure in any serious formulation of general relativity: applications of general relativity to actual physical systems will go on to specify additional beables, including particular fields or particles that contribute to the stress-energy tensor, not to mention the orientation and magnification of telescopes involved in experimental tests of the theory. The set of local beables acknowledged in this way by a theory includes but is almost always much larger than the class of beables involved in formulating it. The settings of switches and knobs on experimental equipment, the currents in coils, and the readings of instruments may be involved in experimental applications of quantum theory, but only an extreme operationalist would argue that these local beables are required for its formulation.

I shall call any beables figuring in a formulation of a theory fundamental, and any additional beables involved in applications of that theory non-fundamental. It follows that, unless they also figure in a formulation of quantum theory, the apparatus of "observation", or better, of experimentation, and the experimental results, are at best non-fundamental beables. Quantum theory acknowledges these beables in some (though not all) applications, but its formulation need not (and, I maintain, should not) involve them. For the moment this leaves the question of the fundamental beables of quantum theory, if any, open.

Based on his intuitive principle of local causality, Bell formulates the following condition on theories:

> *Local Causality*
> A theory is said to be locally causal if the probabilities attached to values of local beables in a space-time region 1 are unaltered by specification of values of local beables in a space-like separated region 2, when what happens in the backward light cone of 1 is already sufficiently specified, for example by a full specification of all local beables in a space-time region 3. ([2004], pp.239-40)

The regions 1-3 are as indicated in figure 1. Several points here require comment and clarification.

Any application of Bell's *Local Causality* condition requires a prior specification of the relevant classes of local beables. But note that these need not be the fundamental beables of the theory to which it is to be applied: at least some may be non-fundamental beables, acknowledged by the theory in applications but not mentioned in formulating it. Indeed, in the intended application of this condition the local beables in regions 1 and 2 may be taken to be the readings and settings of instruments such as photon (atom, or other system) detectors. It may turn out to be possible to apply this *Local Causality* condition to a theory without identifying local beables in that theory—the theory's fundamental beables.

Even given a specification of local beables in regions 1 and 2, Bell's *Local Causality* condition is applicable only to theories that attach a well-defined probability to each measurable set of values of one or more local beables in region 1. This raises a number of issues.

The condition is not applicable to a theory that permits local beables in region 1 to assume certain values while disallowing others, without assigning any probabilities. But no such theory is of interest in the intended applications, since here it would be too weak to predict statistics successfully predicted by quantum theory.

A more important issue concerns the moment at which a theory is presumed to attach probabilities to values of local beables in region 1. In a relativistic space-time, the natural notion of a moment of time is a space-like Cauchy surface: such a moment may be said to be before (events in) region 1 just in case it bisects the backward light-cone of 1 (though not 1 itself), but after (events in) region 1 just in case it bisects the forward light-cone of 1 (though not 1 itself). Only a deterministic theory can be expected to attach the same probability (viz. 0 or 1) to some value of a local beable in region 1 at every moment: a probabilistic theory may be expected to attach the same probability (either 0 or 1) at every moment after region 1, but typically some intermediate probability at moments before region 1. Such a theory may well attach different probabilities to a value of a local beable in region 1 at different moments before region 1. A precise statement of Bell's *Local Causality* condition requires a specification of when (at what moment before region 1) the theory attaches probabilities to values of local beables in space-time region 1.

Two space-like Cauchy surfaces $t,t^*$ before region 1 may coincide everywhere within the backward light cone of region 1 (on a closed hypersurface $\sigma$), where $t$ but not $t^*$ is before region 2—indeed $t^*$ may be after 2, including a hypersurface $\tau$ closed by the forward light cone of 2 (see figure 2). Bell clearly intended his condition of *Local Causality* as a natural generalization of a condition of local determinism modeled on a theory like Maxwellian electromagnetism (cf. Bell ([2004], pp.53-4). What made the determinism of Maxwellian electromagnetism local was that, according to that theory, the electric and magnetic fields in a region like 1 are wholly determined by their values at places within the backward light cone of 1 at a moment before 1. So when Bell formulated his condition of *Local Causality* he may have taken it for granted that the theories to which it was to apply would attach the same probabilities to values of local beables in space-time region 1 at any pair of moments like $t,t^*$ that share a common part such as $\sigma$ within the backward light cone of region 1.

*Local Causality* makes a deeper implicit assumption about how a theory attaches probabilities to values of local beables in a space-time region 1—the assumption that when a theory attaches a probability at a moment to each measurable set of values of a local beable that probability is unique. Only with this assumption can one speak of *altering* probabilities attached to values of local beables rather than attaching *additional* probabilities to them. To see that this assumption does not go without saying, consider a probabilistic theory like classical statistical mechanics. When applying the theory to an isolated macroscopic system one may consistently attach two different probabilities at $t_1$ to a measurable subset of values of some macroscopic local beables at a later time $t_2$. Since its complete microscopic state is assumed to evolve deterministically, either probability 0 or probability 1 may be attached to that subset: but not knowing the complete microscopic state at $t_1$ one does not know which. The theory may also attach a probability at $t_1$ other than 0 or 1 to this same set of values at $t_2$. This is the operative probability based on the accessible information as encapsulated in a macroscopic description of the system at $t_1$. In such a situation the theory consistently attaches more than one probability at the same moment to a measurable set of values of a local beable at a later time.

These are instances of the general phenomenon that probability assignments to an event are relative to a reference class, so the same event may receive multiple probability assignments, each relative to a different reference class. Even if a theory specifies a finer reference class for attaching a probability at some moment it may imply that information about this reference class is inaccessible at that moment and that the accessible information defines a reference class for a different, and indeed more practically relevant, probability. Quantum theory is such a theory, or so I shall argue in the next section.

Bell's condition of *Local Causality* has been carefully analyzed by Norsen [2011], whose analysis has been further improved by Seevinck and Uffink [2011].

They have focused in their analyses on what exactly is involved in a sufficient specification of what happens in the backward light cone of 1.

This specification could fail to be sufficient through failing to mention local beables in 3 correlated with local beables in 2 through a joint correlation with local beables in the overlap of the backward light cones of 1 and 2. A violation of a local causality condition that did not require such a sufficient specification would pose no threat to the intuitive principle of local causality: beables in the overlap of the backward light cones could be considered the common cause of correlated beables in 1,2. On the other hand, requiring a specification of all local beables in 3, may render the *Local Causality* condition inapplicable in attempting to show how theories meeting it predict correlations different from those successfully predicted by quantum theory.

To see the problem, consider the set-up for the intended application depicted in figure 3. $A,B$ are macroscopic events[1], each usually referred to as the detection of a photon linearly polarized either along or at right-angles to an $a$- or $b$-axis respectively: $a,b$ are events at which each axis is selected. The region previously labeled 3 has been relabeled as 3a, a matching region 3b has been added in the backward light cone of 2, and '3' now labels the entire continuous "stack" of space-like hypersurfaces right across the backward light cones of 1 and 2, shielding off these light cones' overlap from 1,2 themselves. Note that each of 1,$a$ is space-like separated from each of 2,$b$.

In some theories, a complete specification of local beables in 3 would constrain (or even determine) the selection events $a,b$. But in the intended application $a,b$ must be treated as free variables in the following sense: in applying a theory to a scenario of the relevant kind each of $a,b$ is to be specifiable independently in a theoretical model, and both are taken to be specifiable independently of a specification of local beables in region 3. Since this may exclude some *complete* theoretical specifications of beables in region 3 it is best not to require such completeness. Instead, one should say exactly what it is for a specification to be sufficient.

Seevinck and Uffink [2011] clarify this notion of sufficiency as a combination of functional and statistical sufficiency, rendering the label $b$ and random variable $B$ (respectively) redundant for predicting $P_{a,b}(A|B,\lambda)$, the probability a theory specifies for beable $A$ representing the outcome recorded in region 1 given beables $a,b$ representing the free choices of what the apparatus settings are in sub-regions of 1,2 respectively, conditional on outcome $B$ in region 2 and beable specification $\lambda$ in region 3. This implies

$$P_{a,b}(A|B,\lambda) = P_a(A|\lambda) \tag{1a}$$

By symmetry, interchanging '1' with '2', '$A$' with '$B$' and '$a$' with 'b' implies

$$P_{a,b}(B|A,\lambda) = P_b(B|\lambda) \tag{1b}$$

---

[1] To simplify notation, I use each of '$A$','$B$' to denote a random variable, a *value* of that variable, or an event in which that variable acquires a value, trusting that the context will make it clear which is intended.

Seevinck and Uffink [2011] offer equations (1a) and (1b) as their mathematically sharp and clean (re)formulation of the condition of local causality. Together, these equations imply the condition

$$P_{a,b}(A, B|\lambda) = P_a(A|\lambda) \times P_b(B|\lambda) \qquad (2)$$

used to derive CHSH inequalities. Experimental evidence that these inequalities are violated by the observed correlations in just the way quantum theory leads one to expect may then be taken to disconfirm Bell's intuitive causality principle.

In more detail, Seevinck and Uffink [2011] claim that orthodox quantum mechanics violates the statistical sufficiency conditions (commonly known as Outcome Independence, following Shimony)

$$P_{a,b}(A|B, \lambda) = P_{a,b}(A|\lambda) \qquad (3a)$$
$$P_{a,b}(B|A, \lambda) = P_{a,b}(B|\lambda) \qquad (3b)$$

while conforming to the functional sufficiency conditions (commonly known as Parameter Independence, following Shimony)

$$P_{a,b}(A|\lambda) = P_a(A|\lambda) \qquad (4a)$$
$$P_{a,b}(B|\lambda) = P_b(B|\lambda) \qquad (4b)$$

Statistical sufficiency is a condition employed by statisticians in situations where considerations of locality and causality simply don't arise. But in this application the failure of quantum theory to provide a specification of beables in region 3 such that the outcome B is always redundant for determining the probability of outcome A (and similarly with 'A', 'B' interchanged) has clear connections to local causality, as Seevinck and Uffink's [2011] analysis has shown.

In the light of Seevinck and Uffink's [2011] analysis, perhaps Bell's *Local Causality* condition should be reformulated as follows:

> *Local Causality$_{SU}$*    A theory is said to be locally causal$_{SU}$ if it acknowledges a class $R_\lambda$ of beables $\lambda$ in space-time region 3 whose values may be attached independently of the choice of *a,b* and are then sufficient to render *b* functionally redundant and *B* statistically redundant for the task of specifying the probability of *A* in region 1.

The notions of statistical and functional redundancy appealed to here are as follows:

> For $\lambda \varepsilon R_\lambda$, $\lambda$ renders $B$ statistically redundant for the task of specifying the probability of $A$ iff $P_{a,b}(A|B, \lambda) = P_{a,b}(A|\lambda)$.

> For $\lambda \varepsilon R_\lambda$, $\lambda$ renders $b$ functionally redundant for the task of specifying the probability of $A$ iff $P_{a,b}(A|\lambda) = P_a(A|\lambda)$.

Note that this reformulation still
i) assumes that a theory specifies a unique probability of $A$ in region 1, but
ii) does not say at what moment $A$ has that probability.

The claim that quantum theory is not locally causal depends on a comparison between quantum probabilities and the probabilities of equations (2)-(4). Specifically, quantum theory is alleged to violate (2) and (3) but not (4). To evaluate this claim we need to know how to understand $\lambda$ for quantum theory.

The claim is most plausible if one takes $\lambda$ to be specified by an appropriately entangled quantum state such as $\Phi^+ = 1/\sqrt{2}(|HH\rangle + |VV\rangle)$. Application of the Born rule of quantum theory to this state at a moment after 3 but before both 1 and 2 yields values

$$P_{a,a}(A(1)|\lambda) = P_{a\bullet}(A(1)|\lambda) =$$
$$P_{\bullet a}(A(2)|\lambda) = 1/2 = P_{a,a}(A(1), A(2)|\lambda)$$

Hence $\quad P_{a,a}(A(1), A(2)|\lambda) \neq P_{a\bullet}(A(1)|\lambda) \times P_{\bullet a}(A(2)|\lambda)$
so (2) fails, and $\quad P_{a,a}(A(1)|A(2), \lambda) \neq P_{a,a}(A(1)|\lambda) \quad$ so (3) fails.
(Here $P_{a\bullet}(A(1)|\lambda)$ is the probability of $A$ in 1, while $P_{\bullet a}(A(2)|\lambda)$ is the probability of an event of the same type in 2, with the relevant detector set to $a$ in each case. The probability $P_{a,a}(A(1)|A(2), \lambda)$ has been calculated in accordance with the standard definition of conditional probability $P_{a,b}(A|B, \lambda) = P_{a,b}(A, B|\lambda) \div P_{a,b}(B|\lambda)$: this is uncontroversial in quantum theory for compatible $A,B$.)

## 3 Quantum States and Born Probabilities

In a pragmatist view of quantum theory I have advocated elsewhere (Healey[2012a,b, 2013]) a quantum state functions not to represent some novel physical structure but to offer authoritative advice to any agent on the significance and credibility of magnitude claims of the form $M_\Delta(s)$ : The value of $M$ on $s$ lies in $\Delta$, where $M$ is a dynamical variable, $s$ is a physical system and $\Delta$ is a Borel set of real numbers. A quantum state assignment is objectively true (or false): in that deflationary sense a quantum state is objectively real. But its function is not to represent some new quantum beable but to help an agent applying quantum theory to predict and explain what happens, as described in non-quantum language. The truth-value of a quantum state assignment depends on physical conditions specifable in non-quantum terms (i.e. with no mention of quantum states, observables, Born probabilities or other characteristic elements of a quantum-theoretic model). Such a specification is given by a set of significant magnitude claims.[2]

It is the specified conditions that make a quantum state assignment true, when it is: the quantum state is backed by (supervenes on) these conditions—it is not caused by them. The quantum state is perhaps best thought of as offering a useful summary of its backing conditions. The epistemic *function* of the quantum state must be clearly distinguished from its non-epistemic, physical grounds. An agent may be warranted in assigning a quantum state even while ignorant of exactly what physical conditions back that assignment. An assignment may be justified restrospectively by appeal to the success of predictions made on its basis. It is because it is often difficult fully to articulate the specific physical conditions backing a quantum state assignment that the application of quantum theory requires skill extending beyond the ability to manipulate the quantum formalism. But the absence of precise rules for applying it in no way distinguishes quantum theory from classical physics. A precise formulation of quantum theory may be given without using 'measurement' or any other of Bell's proscribed words to formulate the Born rule or any physical "collapse" postulate. It does not detract from this precision that agents applying the theory must assign quantum states to systems without being able precisely to specify the physical conditions backing that application.

---

[2]This may seem circular since it is one function of a quantum state to advise an agent on the significance of such claims (see Healey[2012b]). But to correctly assign a quantum state an agent need not be able to make significant magnitude claims specifying the physical conditions on which it depends.

Quantum states are relational on this interpretation. The primary function of Born probabilities is to offer an agent authoritative advice on how to apportion degrees of belief concerning significant magnitude claims which the agent is not currently in a position to check. It follows that a system does not have a unique quantum state. For when agents (actually or merely hypothetically) occupy relevantly different physical situations they should assign different quantum states to one and the same system, even though these different quantum states are perfectly objective. Each will consequently assign different Born probabilities to a single claim $M_\Delta(s)$ about a system $s$ in a given situation. These different probabilities will then be equally objective and equally correct. This feature of the interpretation will prove crucial in what follows.

*Local Causality* presupposes that quantum theory attaches probabilities to values of local beables in regions 1,2. But quantum theory is not a stochastic theory which at each moment attaches a unique probability to a future event. The Born probabilities it supplies are for physically situated agents to adjust their credences in magnitude claims whose truth values they are not in a position to determine. Though objective, Born probabilities are not beables postulated by quantum theory, intended to specify physical chances. Their function is to guide the beliefs of users of the theory, not to represent physical reality. It is for an agent applying the theory to attach probabilities by applying the Born rule to a quantum state appropriate to that agent's physical (and specifically spatiotemporal) location. Guidance is required only because there are physical limits on the information that is accessible from an agent's physical situation. Now the structure of relativistic space-time imposes strict limits on accessible information, assuming physical processes propagate only within the forward light cone. This means that space-like as well as time-like separated agents face different informational limitations, and so require guidance tailored to their different physical situations.

The relational nature of quantum states and Born probabilities raises a problem in applying Bell's *Local Causality* condition to quantum theory. Bell and others apparently understand (1a) to contain two expressions, each intended to represent a single local magnitude—the probability of $A$. Quantum theory does not attach a unique probability to each value of a local beable that it acknowledges in a later space-time region. We saw in section 2 that it is necessary to specify when (at what moment before 1) a theory is supposed to attach a probability to $A$ in (1a). We have now seen that quantum theory attaches a probability relative to the physical situation of a hypothetical agent of limited capacity, in need of advice before forming credences about selected magnitude claims. An idealized representation of that agent's situation in relativity is by a segment of a time-like world-line, along which the agent's situation changes with the passage of proper time. So quantum theory attaches probabilities not in the first instance at moments of time, but at space-time points.

In many circumstances this will be a distinction without a difference. If $p_1$, $p_2$ are distinct points on $\sigma$ (see figure 2), one can formally distinguish the probability attached at $p_1$ of some local beable $A$ in region 1 from its probability attached at $p_2$: but the values of these magnitudes will be equal in this scenario. Moreover, each of these probabilities is independent of whether one considers $\sigma$ part of moment $t$ or of moment $t^*$. But the probability of $A$ attached at $p$ ($p\varepsilon\sigma$) is not the same magnitude as the probability of $A$ attached at $q$ ($q\varepsilon\tau$), and there are applications of quantum theory in which these magnitudes have different values.

One such application is to the EPR-Bell scenario depicted in figure 3, where $a,b$ are now assumed to occur later than $t$. Consider time-like world-lines of two hypothetical agents, Alice and Bob: Alice's world line is confined to the light cone of 1, while Bob's is confined to the light cone of 2. Suppose that physical conditions in the overlap of their backward light cones determine that Alice and Bob would be correct to assign a quantum state to each pair of emitted photons as follows:
i) at every moment on Alice's world-line prior to 1 the pair is assigned polarization state $\Phi^+$, and
ii) at every moment on Bob's world-line prior to 2 the pair is assigned polarization state $\Phi^+$.
Consider a class of pairs in which the outcome of a measurement of photon polarization in 2 is $B$ (i.e. $R$-photon recorded as polarized parallel to the $b$ axis). This outcome backs the assignment at any point on Bob's world-line later than 2 but not 1 of polarization state $|B\rangle$ to the corresponding $L$-photons whose polarization is recorded in 2: the consequent Born probability for such a photon to be recorded as having polarization parallel to the $a$-axis is therefore $P_a^{|B\rangle}(A) = |\langle A|B\rangle|^2$. This is the probability quantum theory attaches at $q$ to that record. By contrast, the probability $P_{a,b}^{\Phi^+}(A) = P_a^{\Phi^+}(A)$ quantum theory attaches at point $p$ to that same record is given by application of the Born rule to quantum state $\Phi^+$, namely $\frac{1}{2}$. Only when the angle between $a,b$ is $\pi/4$ does quantum theory attach a unique value to "the" probability at $t^*$ for an $L$-photon in this class to be recorded as having polarization parallel to the $a$-axis.

Quantum theory does attach a unique value ($\frac{1}{2}$) to the probability $P_{a,b}^{\Phi^+}(A) = P_a^{\Phi^+}(A)$ at $t$ for an $L$-photon to be recorded as having polarization parallel to the $a$-axis. The probability at each point of $t$ is calculated by applying the Born rule to the quantum state $\Phi^+$ correctly assigned at that point. But quantum theory also attaches a unique value at $t$ to the *conditional* probability $P_{a,b}^{\Phi^+}(A|B) = |\langle A|B\rangle|^2$, which equals $\frac{1}{2}$ only if the angle between $a,b$ is $\pi/4$. The standard way of understanding a conditional probability like $P_{a,b}^{\Phi^+}(A|B)$ is through the equation

$$P_{a,b}^{\Phi^+}(A|B) = \frac{P_{a,b}^{\Phi^+}(A,B)}{P_{a,b}^{\Phi^+}(B)} \tag{5a}$$

often taken as a definition of conditional probability, and still well defined in quantum theory since $A,B$ are compatible (they are represented by commuting projection operators). Here the joint probability $P_{a,b}^{\Phi^+}(A,B)$ follows from application of a version of the Born rule generalized to apply to such a set of compatible properties.

The function of a conditional probability magnitude like that on the left of (5a) is to guide an agent's conditional credence: her degree of belief in $A$, assuming $B$. This contrasts with the function of the probability magnitude $P_{a,b}^{\Phi^+}(A) = P_a^{\Phi^+}(A)$, which is simply to guide an agent's credence in $A$. These are two different magnitudes, with two different functions. At $t$ each of Alice and Bob should have credence $\frac{1}{2}$ in $A$, but conditional credence $|\langle A|B\rangle|^2$ in $A$, assuming $B$. Quantum theory, in this pragmatist view, denies that there is any single probability of $A$ at $t$.

Where does this leave *Local Causality*? Note first that in this pragmatist view, quantum theory acknowledges no relevant beables in region 3 of figure 3: the state $\Phi^+$ is not a beable, and the conditions that back it are not present in region 3. So conditionalizing on $\lambda$ in equations such as $(1)-(4)$ is at best vacuous for quantum theory.

Apparently assuming the uniqueness of each of "the probabilities attached to values of local beables in a space-time region 1", Bell's ([2004], pp. 239-40) formulation imposes the condition that this not be *altered* "by specification of values of local beables in a space-like separated region 2...". Read in this way, quantum theory would fail to satisfy this condition if there were a unique probability magnitude attached to each (measurable set of) values of some local beable in 1 for which it offered two distinct estimates, where the estimate neglecting values of local beables in 2 is altered, and typically improved, by a second estimate that specifies these values. But we have seen that in this view of quantum theory it is not the function of Born probabilities to provide estimates of any such unique probability magnitude, either at $t$ or at $t^*$. So on this reading quantum theory cannot fail to be a locally causal theory—not because it satisfies *Local Causality* but because that condition is simply inapplicable to quantum theory. The same analysis applies to *Local Causality$_{SU}$*. The conditions of statistical and functional redundancy both assume uniqueness of *the* probability of $A$ in region 1—presumably the probability $P_{a,b}^{\Phi^+}(A|B)$—for whose specification $b$ but not $B$ proves redundant.

One may object that this conclusion rests entirely too much weight on Bell's choice of the word 'altered' in his formulation of *Local Causality*. Indeed, he did not use this word in his first published formulation of a local causality condition ([2004], p.54). To some this may suggest an alternative reading of *Local Causality* which merely requires in this scenario that the values of $P_{a,b}^{\Phi^+}(A|B)$, $P_a^{\Phi^+}(A)$ be equal even when it is allowed that these Born probabilities do not have the function in quantum theory of providing alternative estimates for a single magnitude—*the* probability of $A$ in region 1. Read in this way, Bell's condition of *Local Causality* is *not* satisfied by quantum theory in this pragmatist view.

But even when Bell ([2004], p.54) first formulated a local causality condition, he introduced it as a natural generalization of a condition of local determinism to stochastic theories that (at each moment) uniquely specify not the subsequent *values* of local beables but their *probability distributions*. In such a theory $P_{a,b}^{\Phi^+}(A|B)$, $P_a^{\Phi^+}(A)$ must be regarded as alternative theoretical specifications of a unique probability distribution, differing (if at all) only when the former provides the better estimate since it is based on additional (surprisingly) relevant information. That is why I do not find this alternative reading plausible. But in the end it is unimportant whether one understands the condition of *Local Causality* to be violated by quantum theory (including relativistic quantum field theory) or simply inapplicable to that theory. As the next section will show, on neither understanding does quantum theory conflict with the intuitive principle on which Bell based that condition.

# 4 Counterfactuals and Causation

Suppose Alice and Bob agreed that when far apart (space-like separated) each would measure the polarization of a different photon from an entangled pair in state $\Phi^+$ along an axis selected randomly at $a$ for Alice, $b$ for Bob. They have repeated this many times on many pairs and amassed robust statistics of their outcomes. While the statistics in the data depend counterfactually on $\Phi^+$, this dependence is not physical since $\Phi^+$ does not describe the condition of the photon pair, in this pragmatist view. But other conditions in the physical world make this the right polarization quantum state for Alice and Bob to ascribe to their photon pairs. These conditions may be expressed by true, significant magnitude claims about physical systems involved in the emission of the pairs. The statistical patterns in the data physically depend on these backing conditions. Different conditions, backing a different polarization state $\Psi$, would almost certainly have resulted in different patterns of outcomes.

The statistics collected by Alice and Bob display striking correlations. While Alice's relative frequency of parallel and orthogonal outcomes is independent of $b$ and Bob's relative frequency of parallel and orthogonal outcomes is independent of $a$, the relative frequency of Alice's parallel and orthogonal outcomes is not generally independent of Bob's outcome for fixed $b$. Indeed, if one idealizes their results, then if $b = a$ there is a perfect correlation between Alice's and Bob's outcomes—her photon's polarization is recorded as parallel to this axis if and only if his is. All these patterns of correlation are correctly predicted by quantum theory. The statistical patterns in their data are just what anyone accepting quantum theory should expect on the basis of the Born probability distributions $P_{a,b}^{\Phi^+}(A, B)$ for state $\Phi^+$ (which each of them correctly ascribed to each pair prior to his or her measurement).

By itself this is not enough to constitute an explanation of their data, any more than the falling barometer suffices to explain the storm it gives us reason to expect. What it leaves out is an account of what those patterns depend on. The settings $a,b$ depend on whatever physical processes set the orientations of the polarization detectors: The correlations also counterfactually depend on physical conditions backing state $\Phi^+$. When those conditions make $\Phi^+$ the correct state to assign to the photon pairs, these counterfactual dependencies suffice to explain the correlations at settings $a,b$. But for generic $a,b$, $P_{a,b}^{\Phi^+}(A|B) \neq P_a^{\Phi^+}(A)$ and $P_{a,b}^{\Phi^+}(B|A) \neq P_b^{\Phi^+}(B)$ even when those processes and conditions are fixed. This indicates a further mutual counterfactual dependence between the physical outcomes $A,B$. Arguably, Bob can explain Alice's statistics by citing her outcome $A$ as counterfactually dependent on his outcome $B$, while she can explain Bob's statistics by citing his outcome $B$ as counterfactually dependent on her outcome $A$.[3] It is by appeal to such counterfactual dependencies that (actual or merely hypothetical) physically situated agents can explain the observed statistics by applying quantum theory locally to show both that they were to be expected and what they physically depend on.

But the patterns of correlation in Alice and Bob's data successfully predicted by quantum theory seem to cry out for a *causal* explanation[4]. The setting events $a,b$ and events in the overlap of the backward light cones of 1,2 involved in preparation of state $\Phi^+$ are naturally considered causes of these patterns, but they leave unexplained the residual mutual counterfactual dependence of $A,B$. This has convinced many that the observed correlations themselves manifest some kind of space-like causal influence or interaction linking space-like separated events (as are events in regions 1 and 2). Why doesn't the explanation quantum theory offers simply show the world is non-local by revealing the nature of that link? Isn't the counterfactual dependence manifested by the failure of (Factorizability) in state $\Phi^+$ simply *causal* dependence?

$$P_{a,b}(A, B) = P_a(A).P_b(B) \qquad \text{(Factorizability)}$$

Bertrand Russell was famously skeptical about the status of causation in fundamental physical theory, and so are many contemporary philosophers. But there is an influential approach to causation in science that squares nicely with a pragmatist view of quantum theory. To introduce this approach, recall Bell's remark:

---

[3] A companion paper will offer such an argument as well as a possible response. I am presently inclined to endorse the argument and to conclude that these are indeed explanations, however shallow.

[4] When Bell ([2004], p.152) says "The scientific attitude is that correlations cry out for explanation", the context makes clear that this is the kind of explanation he has in mind. He there argues that correlations violating (Factorizability) are "*locally inexplicable. They cannot be explained without action at a distance*" (p.153).

> We cannot know what would have happened if something had been different. ... Physical theories are more amenable in this respect. We can calculate the consequences of changing free elements in a theory, be they only initial conditions, and so explore the causal structure of the theory. ([2004], p.101)

Bell made this remark in the course of a characteristically sensitive investigation of what it means to consider an element of a theory free. He clearly regards initial conditions as freely specifiable when building a certain kind of theoretical model of a physical system. It may be that these initial conditions are also subject to external control by a skilled experimenter who constructs such a system. But even if this is not so (consider a model of galaxy formation) the theoretician may freely choose one set of initial conditions rather than another even after deciding how to model a kind of physical system. For a system modeled by deterministic equations, the theoretician may choose to specify final or intermediate rather than initial conditions. That such a choice seems unnatural is a sign that it is made with a nod to the experimentalist or other agent, whose choices are more limited in this respect.

Bell is particularly concerned in this article to establish the freedom of polarizer settings in an EPR-Bell scenario. He notes that from the theoretician's point of view, whether such an element is considered free may depend on what he has decided to model: an element considered free when modeling one system will no longer count as free when that system is expanded to include these elements in the model. Here an element is free if it is exogenous to the system being modeled: Bell's examples are external fields and sources. But, as he notes, these

> are invoked to represent experimental conditions. They also provide a point of leverage for "free-willed experimenters". ([2004], p.101)

Again the theoretician's freedom is influenced if not constrained by what kinds of interventions in the system are available to an experimenter. This is where causal considerations become relevant to theoretical models. When exploring the consequences of changing exogenous elements of a model the theoretician naturally focuses on those elements he views as potential levers for intervention, even when no such intervention is contemplated or even physically possible. These are the elements whose variation enables him to explore the causal structure of the theory. At least for a deterministic model, they include initial, but not final, conditions.

Interventionist approaches to causation are now prominent in the medical and social sciences, and their philosophical foundations have been extensively explored. They are motivated not by the desire to analyze the causal relation in counterfactual or other terms but to systematize our understanding of the interplay between prior causal assumptions and theoretical (especially probabilistic) scientific models. The causal structure of a model is to be investigated by considering the consequences of hypothetical external interventions, themselves characterized in explicitly causal terms. Quantum theory supplies us with probabilistic models of systems that display "non-local" correlations, so we can explore their causal structure by isolating points where intervention is possible and seeing what would follow from such intervention.

Consider how quantum theory models a 2-photon system in state $\Phi^+$ described in section 2 in the scenario depicted in figure 3. $\Phi^+$ is the initial (polarization) state of the system, and Bell may well have regarded it as a free variable, subject to external intervention. One who denies that the quantum state is a beable must look elsewhere for physical initial conditions—to the claims that back this state assignment. These are indeed subject to external intervention: an experimenter could have prepared a different quantum state $\Psi$. Some such interventions would have altered the Born probabilities $P_a^{\Phi^+}(A)$, $P_b^{\Phi^+}(B)$ of outcomes $A,B$ in 1, 2: so some event described by a claim backing state assignment $\Phi^+$ counts as a probabilistic common cause of those outcomes.

The polarizer settings are free variables in the model: events $a$, $b$ are within an experimenter's control, which he may, for example, delegate to a quantum random number generator. So too are the relevant aspects of the initial state of each detector and its environment, though there will be no need to consider interventions on these: it is necessary only to assume that they are adjusted so as to record coincident outcomes in 1, 2 often enough to collect extensive records to compare against predicted Born probabilities.

For each assignment of initial quantum state and choice of polarizer settings, quantum theory yields a probability distribution over four "fine-grained" quantum models, each of which includes a different combination of outcomes in 1, 2: these outcomes are omitted in a "coarse-grained" model. Quantum theory functions not to predict a particular combination of outcomes (as specified in a fine-grained model) but to advise a situated agent like Alice and Bob on what statistical distribution of these combinations to expect, given a choice of polarizer settings and assignment of a particular initial quantum polarization state.

This means that the actual outcomes recorded for a pair in 1, 2 are elements in some fine-grained model and so are not exogenous to the quantum modeling structure. They appear neither in the coarse-grained model nor in all the fine-grained models. But the application of quantum theory here presupposes that some outcomes are recorded, and that in each instance of coincident records the unique actual outcome pair is represented in exactly one fine-grained model. Since the values of the variables $A$, $B$ are represented as fixed elements in fine-grained quantum models, they are not exogenous, and so they are not free elements even in the broad sense implicit in the theoretician's use of quantum theory. Nor are they subject to any physically possible intervention.

How could one try to make sense of an intervention on $B$ with respect to $A$ or *vice versa*? Certainly no alternative action of Alice, Bob or other agent could set the value of $A$ or $B$ without disrupting the system involved in the EPR-Bell scenario itself (e.g. by preparing a different quantum state). In particular, choosing to measure a different polarization component would not have this effect. Quantum theory itself provides no resources on which one can draw to make sense of an intervention capable of changing the outcome of Alice's or Bob's measurement of a fixed component of polarization.

If it does not make sense to speak of an intervention in a system capable of making the antecedent of a counterfactual conditional true, then this is not a causal counterfactual. Consider an ideal case in which $b = a$, so that $P_{a,b}^{\Phi^+}(A_{Alice}|A_{Bob}) = 1$. Any agent in a position to assign state $\Phi^+$ should then endorse the counterfactuals (*Same*), (*Same*$^*$):

(*Same*) Had Bob recorded the opposite outcome then Alice would have also.

(*Same*$^*$) Had Alice recorded the opposite outcome then Bob would have also.

But the literature on causal modeling shows how important it is to distinguish between conditioning and intervening when assessing such counterfactuals and the conditional probabilities on which they depend. The probability of $Y$ conditional on the value of $X$ is in general different from the probability that $Y$ would have if an intervention had set $X$ to that value. Only if the probability distribution $P_{a,b}^{\Phi^+}(A_{Alice}|A_{Bob})$ is invariant under a suitable intervention that sets $A_{Bob}$ would (*Same*) express a causal counterfactual.[5] More generally, to use the fact that $P_{a,b}^{\Phi^+}(A|B) \neq P_a^{\Phi^+}(A)$ to argue for a causal connection between $A$ and $B$, one must show that this inequality remains invariant under suitable interventions on $B$. But that issue cannot even be raised unless such an intervention makes sense.

In his sophisticated discussion of what the possibility of intervention requires, Woodward ([2003], pp.130-3) argues that an intervention must be conceptually possible, though it need not be physically possible. He considers a case in which an event $C$ that is a potential locus of intervention occurs spontaneously in the sense that it has no causes. He argues that even in this case one can make sense of an intervention on $C$. This suggests that one might make sense of the idea that an intervention in the EPR-Bell scenario is capable of making true the antecedent of (*Same*) or (*Same*$^*$) in the case under consideration.[6] I cannot see how to do so on the present pragmatist view of quantum theory. But the following argument that (*Same*), (*Same*$^*$) are not causal counterfactuals assumes only that it makes sense to intervene on $B$ if and only if it makes sense to intervene on $A$—a natural assumption in the light of Lorentz invariance.

Either interventions on $B$, $A$ are both (conceptually) possible or neither intervention is possible. If neither intervention is possible, then interventionist conditional probabilities such as $P_{a,b}^{\Phi^+}(A_{Alice}|do[A_{Bob} = x_i])$ are not well-defined, depriving counterfactuals (*Same*), (*Same*$^*$) of a significant causal reading. In that case no-one accepting quantum theory should take either (*Same*) or (*Same*$^*$) to establish a causal connection between Alice's and Bob's outcomes.

---

[5] Pearl ([2009], p.70) would express such stability in the form $P_{a,b}^{\Phi^+}(A_{Alice}|A_{Bob}) = P_{a,b}^{\Phi^+}(A_{Alice}|do[A_{Bob} = x_i])$, in which $do[A_{Bob} = x_i]$ represents an intervention setting the value of $A_{Bob}$ to $x_i$.

[6] Hausman and Woodward ([1999], [2004]) reject this suggestion, arguing that a distinction between intervening on $X$ with respect to $Y$ and acting directly on both $X$ and $Y$ cannot be drawn in this case. But their argument for this conclusion depends on controversial assumptions not shared by the present pragmatist view of quantum theory.

Instead, suppose interventions on $B, A$ are both (conceptually) possible. Woodward ([2003], p.98) states necessary and sufficient conditions for $I$ to be an intervention variable for $X$ with respect to $Y$. These include

**(I2)**  $I$ acts as a switch for all the other variables that cause $X$. That is, certain values of $I$ are such that when $I$ attains those values, $X$ ceases to depend on the values of other variables that cause $X$ and instead depends only on the value taken by $I$.

Let $X$ be a variable with values 0,1 according as Bob gets outcome $A^{\|}_{Bob}$, $A^{\perp}_{Bob}$ respectively, and $Y$ be a variable with values 0,1 according as Alice gets outcome $A^{\|}_{Alice}, A^{\perp}_{Alice}$ respectively for their space-like separated measurements of polarization along axis $a$ in state $\Phi^+$: and assume $J$ is some hypothetical intervention variable for $X$ with respect to $Y$ here. Anyone accepting quantum theory should believe the counterfactual conditional ($Same$) holds of outcomes in the absence of interventions. It does not follow that any agent accepting quantum theory should believe $X$ causes $Y$, since neither the counterfactual nor the conditional probability on which it depends need remain stable under intervention on $X$ with respect to $Y$, and the physical impossibility of intervening on $X$ leaves it quite unclear how to decide whether it does. Nevertheless, assuming such stability, the (conceptual) possibility of an intervention on $X$ with respect to $Y$ would establish that $X$ is a cause of $Y$.

Applied to ($Same^*$), a parallel argument assuming an intervention on $Y$ with respect to $X$ should convince anyone accepting quantum theory that $Y$ causes the corresponding value of $X$. However if $Y$ is indeed a cause of $X$ then, since $J$ was assumed to act as a switch, **(I2)** implies that $X$ ceases to depend on $Y$, so an intervention on $X$ will not change the value of $Y$. But if intervening by changing Bob's outcome does not change Alice's outcome then the counterfactual dependence ($Same$) does *not* imply a corresponding causal dependence. The assumption that it is possible to intervene by changing Bob's outcome implies that Alice's outcome is causally independent of Bob's. But if it is not possible so to intervene then ($Same$) cannot be given a causal reading. By symmetry of reasoning, nor can ($Same^*$). Therefore the counterfactuals ($Same$), ($Same^*$) are not causal.

The argument for this conclusion assumed a symmetry condition on any (conceptually) possible interventions on outcomes $A,B$ that comports with Lorentz invariance if 1,2 are space-like separated. It does not exclude $A_{Bob}$ as a cause of $A_{Alice}$, or $A_{Alice}$ as a cause of $A_{Bob}$, when these events are time-like separated. But to establish some direct causal relation between time-like separated events such as $A_{Bob}$ and $A_{Alice}$ one would first have to make sense of an intervention on one with respect to the other, and then provide reasons to believe that the conditional probability of one on the other remains invariant under such intervention. I see no reason to believe that can be done. Temporal precedence of cause to effect may help us distinguish cause from effect when we are dealing with a pair of directly causally connected events. But it does not make the earlier of two counterfactually dependent events into a cause of the later if these are not directly causally connected.

On an interventionist approach to causation amenable to a pragmatist in-terpretation of quantum theory, anyone accepting quantum theory should take some event described by claims backing the assignment of an entangled state such as $\Phi^+$ to be a common cause of the outcome events recording the polarizations of the photons concerned, whatever the space-time intervals relating these outcome events. She should acknowledge that at least given many such common causes the actual setting of each polarizer is also a cause of the outcome recorded by the corresponding detector. But she should reject the claim that a local setting is a cause of a distant outcome, since in this case there is not even any counterfactual dependence between these events. She should reject any claim of causal dependence among the outcomes, whether these are space-like or time-like separated. Finally, she should reject any claim of superluminal influence or interaction. Bell's argument based on *Local Causality* establishes no such thing. If you accept quantum theory you have no reason to believe that our world is non-local in any of these senses.

## 5  Microcausality

After arguing that ordinary quantum mechanics fails to satisfy his condition of *Local Causality* and cannot be embedded in a theory that is locally causal in this sense, Bell ([2004], p.245) considers an alternative expression of the fundamental causal structure of theoretical physics—the requirement of "no signalling faster than light". Already in an earlier paper he had advanced an argument that "ordinary quantum field theory" meets this requirement because of "the usual local commutativity condition" (often called microcausality) that observables localized in space-like separated regions commute. As he there noted ([2004], p.60), that argument rested on assumptions about what *we* can do. It assumed that by acting in the part of the backward light cone of 1 that does not overlap with that of 2 an agent can control the experimental setting in 1 (and specifically the polarization axis $a$) but not the outcome $A$, while by acting in the part of the backward light cone of 2 that does not overlap with that of 1 an agent can control the experimental setting in 2 (and specifically the polarization axis $b$) but not the outcome $B$: and it assumed that we can change the Hamiltonian by changing some external fields.

> ...if the ordinary quantum field theory is embedded in this way in a theory of beables, it implies that faster than light signalling is not possible. In this *human* sense relativistic quantum mechanics *is* locally causal. ... You may feel that only [this] 'human' version [of local causality] is sensible and may see some way to make it more precise. ([2004], p.61)

But he found this hard to accept

> For one thing, we have lost the idea that the correlations can be explained, or at least this idea awaits reformulation. More importantly the 'no-signalling...' notion rests on concepts which are desperately vague, or vaguely applicable. The assertion that 'we cannot signal faster than light' immediately provokes the question: 'Who do we think *we* are?' ([2004], p.245)

So Bell draws two conclusions:
(i) Despite its connection to a no-superluminal-signalling requirement, microcausality is not justified as an expression or requirement of "the fundamental causal structure of theoretical physics".
(ii)The failure of ordinary quantum mechanics to be locally causal (or even embeddable in a locally causal theory) leaves us with no clear idea of to how to explain correlations violating Bell inequalities.

My response to (ii) here must be brief. Section 4 indicated how quantum theory, in a pragmatist view, helps physically situated agents like us to explain correlations violating Bell inequalities by showing that these were to be expected and what they physically depend on. I submit that this does give us a clear idea of how to explain the puzzling correlations even though the explanation cannot be understood in terms of the operation of "near by" (spatiotemporally contiguous) or spacelike separated causes.

In response to (i) I shall argue that:
(a) A strengthened form of microcausality can be justified as a sufficient condition for the consistent assignment of quantum states in Lorentz invariant ordinary quantum theory:
(b) While a no-superluminal-signalling requirement does not express the fundamental causal structure of theoretical physics it does figure in an argument that violation of Bell inequalities is no manifestation of superluminal causation:
(c) Theoretical physics cannot be expected to express any fundamental causal structure, since causal structure depends on the possibility of intervention, and (like 'cause' and 'effect') 'intervention' is not a term or notion that has any place in a precise formulation of physical theory.

To motivate (a) recall that in a pragmatist view a quantum state is assigned at the physical situation of an actual or hypothetical agent, and that an idealized representation of such an agent is by a timelike world-line. Consider the situation of such an agent Charlie within the overlap of the forward light-cones of regions 1, 2 in the scenario depicted in figure 4. After recording polarization $B_R$ for his photon, Bob should assign polarization state $|B_L\rangle$ to Alice's photon in the region of his forward light cone that does not overlap Alice's. If his recording of the linear polarization of photon R in 2 were to proceed by an ideal projective measurement, then in that same region he should assign polarization state $|B_R\rangle$ to photon R, and state $|B_L\rangle \otimes |B_R\rangle$ to the pair. If Alice's coincident recording of the linear polarization $A_L$ of photon L in 2 were to proceed by an ideal projective measurement, then in the region of her forward light cone that does not overlap Bob's she should assign polarization state $|A_L\rangle$ to photon L, $|A_R\rangle$ to R, and $|A_L\rangle \otimes |A_R\rangle$ to the pair.[7]

---

[7]The motivation does not presuppose that such measurements can actually be carried out. Ideal projective measurements of spin component may prove a more practicable proposition than those for photon polarization.

In the overlap of the forward light cones of 1, 2 Charlie has potential access to Alice's and Bob's records. To determine the correct assignment of polarization state to the pair following Alice and Bob's ideal projective measurements, Charlie may proceed in either of two ways. He may follow Alice in assigning state $|A_L\rangle \otimes |A_R\rangle$ within the region of her forward light cone that does not overlap Bob's and then assign state $|A_L\rangle \otimes |B_R\rangle$ at his location, taking account of the effect and result of Bob's measurement in 2: or he may follow Bob in assigning state $|B_L\rangle \otimes |B_R\rangle$ within the region of Bob's forward light cone that does not overlap Alice's and then assign state $|A_L\rangle \otimes |B_R\rangle$ at his location, taking account of the effect and result of Alice's measurement in 1. In this special case of ideal projective measurements consistency is secured by the fact that both procedures lead to the same state assignment within the overlap of Alice's and Bob's forward light cones.

As in this motivating example, ordinary quantum field theory represents a local operation in a bounded region of spacetime by an operator on a Hilbert space associated with that region. Microcausality is usually understood as the requirement that observables localized in space-like separated regions commute. But, unlike observables, not all local operations are represented by self-adjoint operators. For any local observable $C$ in region 1 represented in ordinary quantum mechanics by a self-adjoint operator on a Hilbert space with pure discrete spectrum, its spectral projections $\{P_i^C\}$ form a set of self-adjoint measurement operators for an ideal projective measurement. So if $D$ is also such a local observable in a space-like separated region 2 with spectral projections $\{P_j^D\}$ then a consistent assignment of state to a system following joint ideal projective measurements of $C$ and $D$ is guaranteed if $\forall i, j \; [P_i^C, P_j^D] = 0$, or, in other words, if $[C, D] = 0$. So if the only local operations permitted are ideal projective measurements of observables represented by self-adjoint operators on a Hilbert space with pure discrete spectrum then the requirement that observables localized in space-like separated regions commute suffices for the consistent assignment of quantum states in Lorentz invariant ordinary quantum field theory. How can such consistency be guaranteed for every type of local operation on the quantum state of a system?

In ordinary quantum mechanics such state operations ("measurements") may be modeled by POVMs, and the state following a POVM $\{E_i\}$ may be specified in terms of a set of measurement operators $\{M_i\}$ compatible with (but not defined by) it:

$$E_i = M_i^\dagger M_i \quad \text{where} \sum_i E_i = I \qquad (6)$$

in which case the appropriate rule for updating a quantum state with density operator $\rho$ following a measurement with $i$th outcome is

$$\rho \longrightarrow \rho' = \frac{M_i \rho M_i^\dagger}{Tr[M_i \rho M_i^\dagger]} \qquad (7)$$

Suppose Alice's measurement is represented by $\{M_i^A\}$, Bob's by $\{M_j^B\}$. Then Charlie's two state assignment procedures applied to a state $\rho$ (such as $|\Phi^+\rangle \langle \Phi^+|$)

yield states

$$\rho^{AB} = \frac{M_j^B M_i^A \rho M_i^{A\dagger} M_j^{B\dagger}}{Tr[M_j^B M_i^A \rho M_i^{A\dagger} M_j^{B\dagger}]} \tag{8}$$

$$\rho^{BA} = \frac{M_i^A M_j^B \rho M_j^{B\dagger} M_i^{A\dagger}}{Tr[M_i^A M_j^B \rho M_j^{B\dagger} M_i^{A\dagger}]} \tag{9}$$

Since $\rho^{AB} = \rho^{BA}$ if $[M_i^A, M_j^B] = 0$, a sufficient condition for consistent state assignments following state operations in space-like separated regions is that the corresponding sets of measurement operators pairwise commute. Since not all measurement operators are self-adjoint, this represents a strengthening of the usual microcausality condition. This strong microcausality condition is motivated (though not entailed) by Lorentz invariance. If it were to fail, a consistent quantum state assignment at a space-time region following a pair of operations in space-like separated regions could presuppose an absolute time order of those operations, in violation of fundamental Lorentz invariance.

I turn now to (b). Even if there were some direct causal connection between events in space-like separated regions it would not follow that this could be used to signal superluminally. There might be no controllable cause variable in either region, and/or it might be that no effect variable in either region is observable. But a no-superluminal-signalling requirement may still figure importantly in an argument that there is no direct causal connection between events in spacelike separated regions.

On an interventionist approach, to establish a causal connection between such regions it is necessary to demonstrate the persistence of a counterfactual connection between values of variables (one in each region) under possible interventions. Superluminal signalling is possible only if a counterfactual connection persists under some such physically possible intervention. So while the impossibility of superluminal signalling does not exclude superluminal causation, it does rebut the simplest and most direct argument for superluminal causation.

Bell ([2004], p.237) considers two classes of physically possible interventions on variables, understood as local beables—on apparatus settings and on values of external fields. He appeals to Eberhard's [1978] no-signalling result to exclude signalling by manipulation of apparatus setting; and to microcausality, formulated as the requirement that the commutator of Heisenberg operators $[\hat{A}(x), \hat{B}(y)]$ representing local observables vanish when evaluated at space-like separated points $x, y$, to exclude signalling by manipulation of external fields. The strong microcausality condition considered above, requiring commutation of measurement operators localized in space-like separated regions, suffices to exclude superluminal signalling by a more general local state operation, however this may be represented as a manipulation of local beables.

These considerations are not just remarks about what we can do. They cover a wide class of physically possible interventions and show that the counterfactual connections between values of local variables in space-like separated regions fail to persist under these interventions. In doing so they provide defeasible evidence against the hypothesis that there is some direct causal connection between events in these regions. To establish such a causal connection in the face of this counterevidence it would be necessary to make sense of a different kind of intervention acting on the value of a local variable in one of two space-like separated regions that preserves its counterfactual connection with the value of a variable in the distant region. The strong microcausality condition implies that a counterfactual connection between distant outcomes is *not* preserved under intervention on any local variable associated with a state transformation. The possibility of a direct intervention on a local outcome was examined and rejected in the previous section. On an interventionist approach to causation suggested by some of Bell's remarks and consonant with a pragmatist view of quantum theory, a no-superluminal-signalling requirement is not just a remark about what *we* can do. By motivating a strong microcausality condition it helps to show that there is no reason to regard violation of Bell inequalities as a manifestation of superluminal causation.

In his conclusion, Bell ([2004], p.245) rejected both *Local Causality* and no-superluminal-signalling as possible expressions of the fundamental causal structure of contemporary theoretical physics—the former because "it does not work in quantum mechanics and this cannot be attributed to the "incompleteness" of that theory", the latter because it is desperately vague. Does its lack of any acceptable expression of fundamental causal structure pose a problem for contemporary theoretical physics? On an interventionist approach to causation it should not.

Bell ([2004], p.215) famously argued for the exclusion of terms including 'measurement','apparatus', 'observable', 'information' from a formulation of a physical theory, even while acknowledging their importance in its application. Along with 'cause', the term 'intervention' belongs on Bell's list of proscribed words. Talk of interventions is appropriate only in actual or hypothetical applications of a theory. The causal structure of any scientific theory is to be explored by applying it to assess what would happen under various hypothetical interventions on its variables. Such interventions involve variables exogenous to the theoretical model under study. Typically, these are not even variables of the theory itself (consider medical interventions in which the intervention variable corresponds to the investigator's action when administering a drug or placebo).

A physical theory may be fundamental in the sense that any model permits extension also to include variables designated as intervention variables. But to evaluate the consequences of intervention *via* those variables one must hold fixed the values of certain other variables (taken to be independent causes of $Y$ in the notation of (**I2**) above). This may be inconsistent with any extension of the original model, making the hypothetical intervention physically (though not conceptually) impossible. Even in a physically possible intervention, the choice as to which other variables to hold fixed depends on causal considerations external to the theory itself. So even a fundamental physical theory cannot by itself specify the consequences of the hypothetical interventions needed to evaluate its causal content. As Bell says, we can explore the causal structure of a physical theory by calculating the consequences of changing free elements.

But this requires an agent's prior decision as to which elements to count as free, and which as fixed, in an application of the theory. Causal structure emerges from fundamental physical theory only from the perspective of agents like us applying that theory. That is why fundamental physical theory itself cannot be expected to express or require such structure.

# 6  Extraordinary Quantum Mechanics

Much recent philosophical discussion of *Local Causality* and microcausality has been in the context of what Ruetsche ([2011]) calls extraordinary quantum mechanics. The quantum mechanics of systems with an infinite number of degrees of freedom implies the existence of unitarily inequivalent representations of the fundamental commutation relations whose rigorous treatment is the topic of algebraic quantum field theory. This provides an appropriate framework within which to investigate the causal structure of relativistic quantum theory. Haag's ([1992]) *Local Quantum Physics* is a classic presentation of the framework. It is advertised as offering a comprehensive account of local quantum physics, understood as the synthesis of quantum theory with the principle of locality. One naturally expects new insights about *Local Causality* and microcausality to emerge from algebraic relativistic quantum field theory. But while many facinating technical questions arise within this framework, answering them seems unlikely to require radical revision of an analysis of the justification and implications of these principles within ordinary quantum mechanics.

In algebraic quantum field theory a quantum state is represented by a complex linear map defined on a C* and/or von Neumann algebra of operators whose value for a self-adjoint operator yields the expectation value of the corresponding observable. The theory is local in that operators and states are defined locally—on open, bounded regions of space-time. A theory is defined by a net of local algebras of bounded operators. In this framework the microcausality condition becomes

> *Algebraic Microcausality*  $[A, B] = 0$ for all $A \in \mathfrak{A}(\mathcal{O}_1)$, $B \in \mathfrak{A}(\mathcal{O}_2)$ with $A, B$ elements of local observable algebras in spacelike separated regions $\mathcal{O}_1, \mathcal{O}_2$ respectively.

Patterns of correlation in violation of Bell inequalities are known to be a quite generic prediction of algebraic relativistic quantum field theory, including those among measurable values of observables localized in space-like separated regions[8]. If *Local Causality* entailed such inequalities (in conjunction with other incontrovertible assumptions) then this would establish the failure of *Local Causality* in algebraic relativistic quantum field theory. But section 4's reasons for questioning the applicability of *Local Causality* to ordinary quantum theory apply equally to algebraic relativistic quantum field theory. If *Local*

---

[8] See, for example, Halvorson and Clifton [2000], reprinted in Butterfield and Halvorson (eds.) [2004], and section 2.1 of the editors' introduction to that volume.

*Causality* is a condition that cannot be applied to this extraordinary quantum theory, then nor can it be said to fail in algebraic relativistic quantum field theory. On the other hand, if *Local Causality* is understood to require merely that correlations among measurable values of observables localized in space-like separated regions obey the Factorizability condition for *every* state, then it fails in algebraic relativistic quantum field theory just as it does in ordinary quantum theory. But the failure of Factorizability for some states in algebraic relativistic quantum field theory is no more untoward than its failure in ordinary quantum mechanics.

Adoption of the algebraic approach to relativistic quantum field theory does involve a minor modification in the significance of microcausality in quantum mechanics. In ordinary quantum mechanics, the condition that operators representing observables localized in space-like separated points or regions commute does not suffice to guarantee a consistent assignment of quantum states in the overlap of their forward light-cones. But, as noted in the last section, a strengthened formulation (also) requiring commutation of measurement operators in such spacelike separated regions does guarantee such consistency. Even if they are not self-adjoint, measurement operators in algebraic quantum field theory may still be elements of local observable algebras. Given that they are, the condition of *Algebraic Microcausality* suffices to guarantee consistent assignment of states in a Lorentz invariant algebraic quantum field theory since it applies to *all* elements of local algebras, not merely to the self-adjoint elements representing observables.

This justification for *Algebraic Microcausality* clarifies its relation to Lorentz invariance. A separate Covariance axiom is often thought to exhaust the content of the requirement that an algebraic quantum field theory be Lorentz invariant by respecting the symmetries of a special relativistic space-time. But even a theory conforming to a standard Covariance axiom could fail to permit state assignments consistent with Lorentz invariance if it did not obey *Algebraic Microcausality*. These considerations do not show *Algebraic Microcausality* to be a necessary condition for Lorentz invariance. Note that $\rho^{AB} = \rho^{BA}$ in equations (8),(9) also if the measurement operators *anti*commute: $\left\{M_i^A, M_j^B\right\} = 0$. This at least suggests a way of ensuring Lorentz invariant state assignments in a theory violating *Algebraic Microcausality*. But imposing *Algebraic Microcausality* certainly makes it easier to guarantee that an algebraic relativistic quantum field theory respect the space-time symmetries required by Lorentz invariance.

# 7 Conclusion

Relativistic quantum theory (in the form of ordinary or algebraic quantum field theory) does not satisfy Bell's *Local Causality* condition. But this does not mean quantum phenomena exhibit superluminal causal interactions. Applied in a pragmatist view of quantum theory, an interventionist approach to causation shows why they don't.

Microcausality is a relativistic requirement not because relativity forbids superluminal signalling but because microcausality ensures that quantum states may be assigned in a relativistically invariant way.

Neither *Local Causality* nor microcausality should have been expected to express the fundamental causal structure of theoretical physics, since causal structure emerges only in a context of possible applications of a theoretical model by hypothetical physically situated agents external to the model.

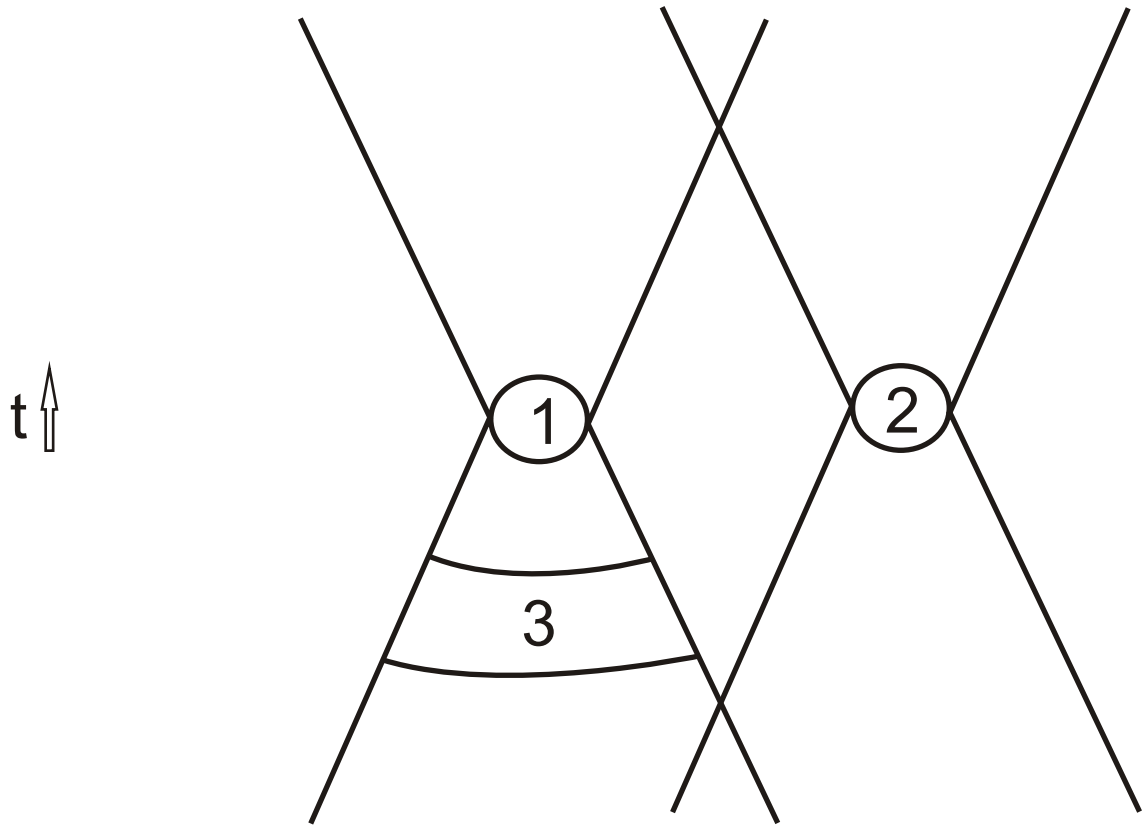

Figure 1

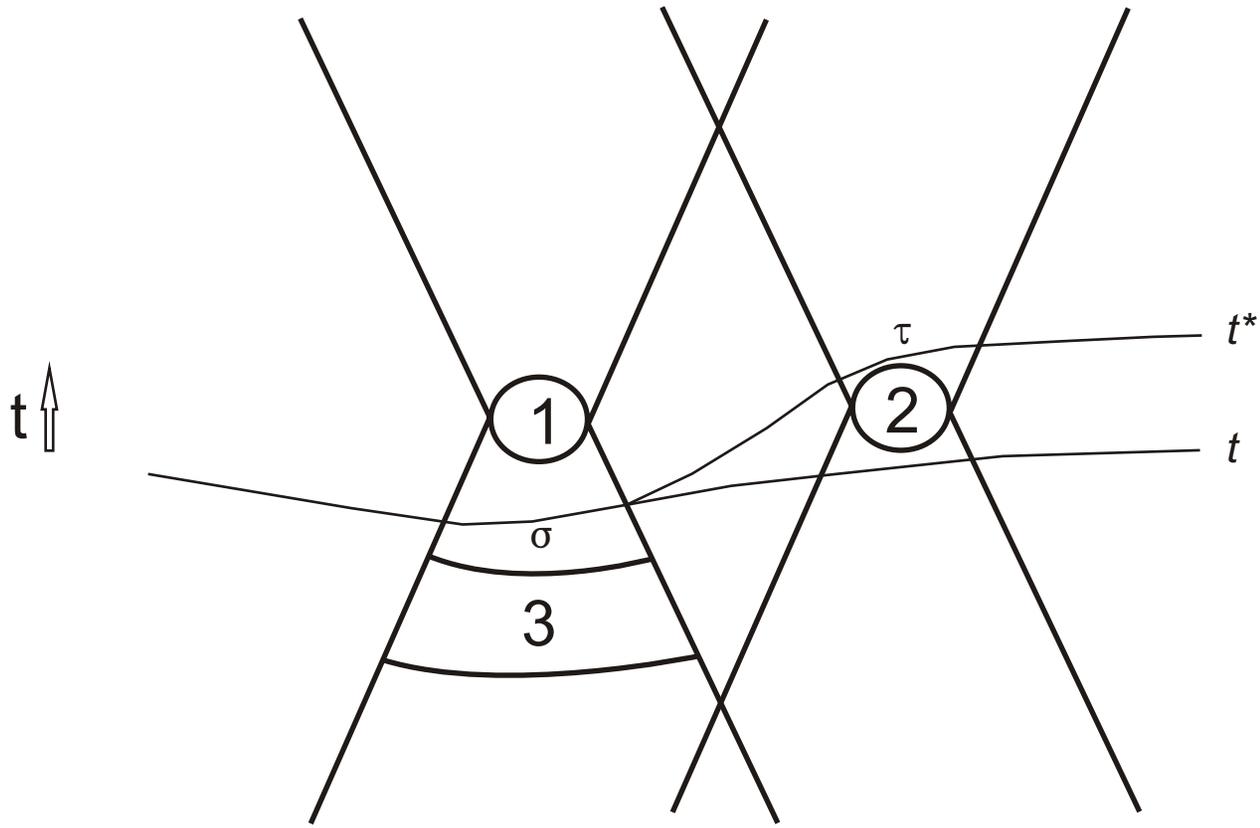

Figure 2

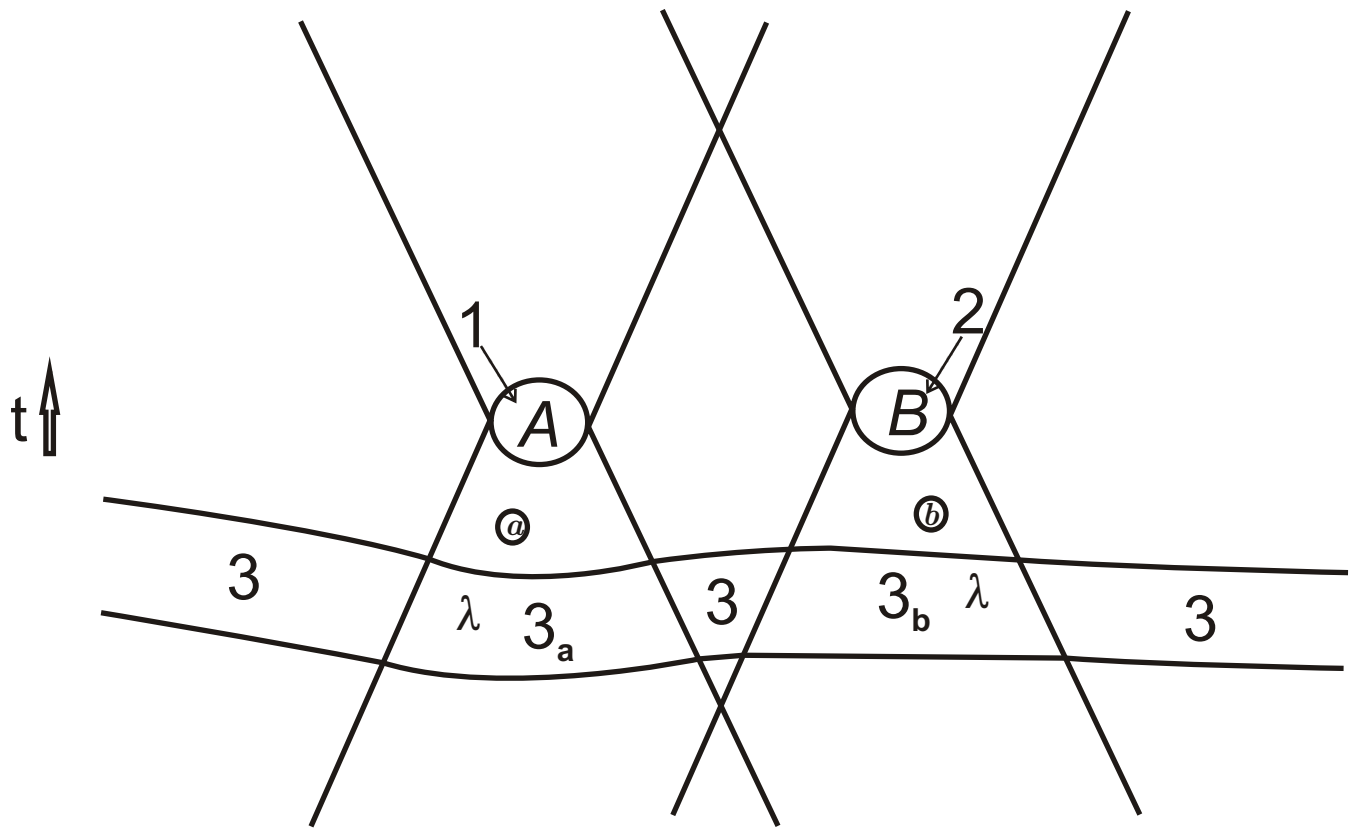

Figure 3

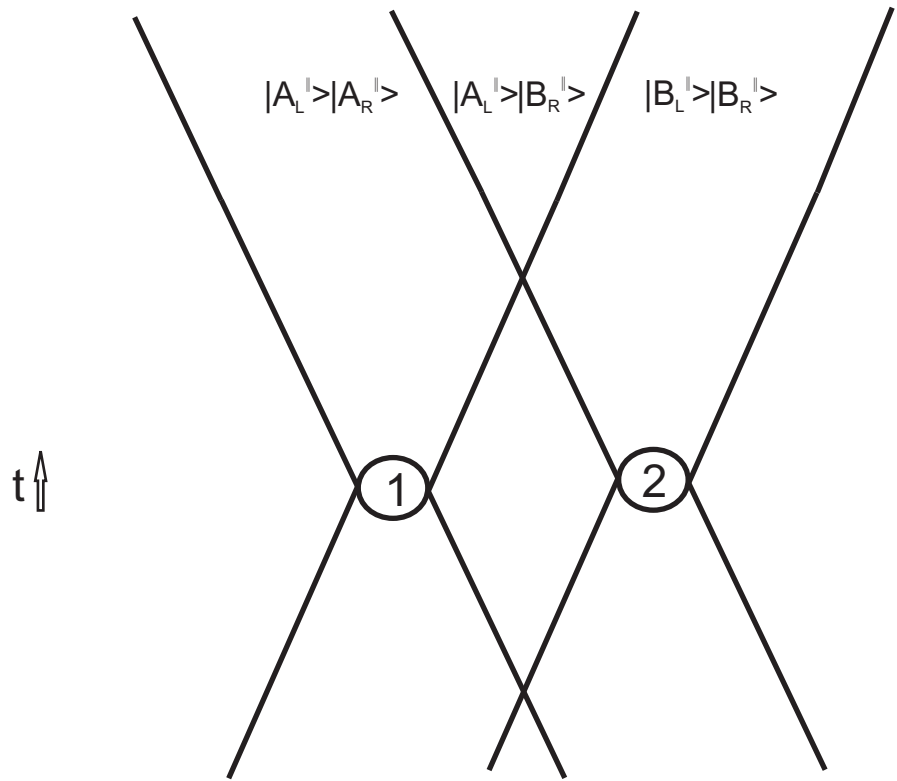

Figure 4